

\font\titolino=cmbx10
\font\tsnorm=cmr10
\font\tscors=cmti10

\font\tscorsp=cmti9
\magnification=1200

\hsize=148truemm
\hoffset=10truemm
\parskip 3truemm plus 1truemm minus 1truemm
\parindent 8truemm
\newcount\notenumber

\def\note{\advance\notenumber by 1 \footnote{$^{\the\notenumber}$}}
\def\ref#1{\medskip\everypar={\hangindent 2\parindent}#1}
\def\beginref{\begingroup
\bigskip
\leftline{\titolino References.}
\nobreak\noindent}
\def\endref{\par\endgroup}
\def\beginsection #1. #2.
{\bigskip
\leftline{\titolino #1. #2.}
\nobreak\noindent}
\def\beginappendix #1.
{\bigskip
\leftline{\titolino Appendix #1.}
\nobreak\noindent}
\def\beginack
{\bigskip
\leftline{\titolino Acknowledgments}
\nobreak\noindent}
\def\begincaptions
{\bigskip
\leftline{\titolino Figure Captions}
\nobreak\noindent}
\nopagenumbers
\rightline{\tscors DFTT 2/93}
\rightline{\tscors January 1993}
\vskip 20truemm
\centerline{\titolino AN ANISOTROPIC WORMHOLE:}
\bigskip
\centerline{\titolino TUNNELLING IN TIME AND SPACE}
\vskip 20truemm
\centerline{\tsnorm Marco Cavagli\`a$^{(a,c)}$,
Vittorio de Alfaro$^{(a)}$ and Fernando de Felice$^{(b)}$}
\bigskip
\centerline{$^{(a)}$\tscorsp Dipartimento di Fisica
Teorica dell'Universit\`a di Torino}
\smallskip
\centerline{\tscorsp and INFN, Sezione di Torino, Italy}
\bigskip
\centerline{$^{(b)}$\tscorsp Istituto di Fisica Matematica
dell'Universit\`a di Torino}
\bigskip
\centerline{$^{(c)}$\tscorsp Present address: ISAS, Trieste}
\smallskip
\vskip 25truemm
\centerline{\tsnorm ABSTRACT}
\begingroup\tsnorm\noindent
We discuss the structure of a gravitational euclidean instanton obtained
through coupling of gravity to electromagnetism. Its topology at fixed
$t$ is $S^1\times S^2$. This euclidean solution can be interpreted as a
tunnelling to a hyperbolic space (baby universe) at $t=0$ or
alternatively as a static wormhole that joins the two asymptotically
flat spaces of a Reissner--Nordstr\"om type solution with $M=0$.
\smallskip\noindent
\smallskip\noindent
\smallskip\noindent
\smallskip\noindent
\smallskip\noindent
\vfill
\hrule
\noindent
Mail Address:
\hfill\break
Dipartimento di Fisica Teorica
\hfill\break
Via Giuria 1, I-10125 Torino
\hfill\break
Electronic mail:VAXTO::VDA or VDA@TORINO.INFN.IT
\endgroup
\vfill
\eject
\footline{\hfill\folio\hfill}
\pageno=1

\beginsection 1. Introduction.
Wormholes (WH) are classical euclidean solutions for the gravitational
field coupled to matter or gauge fields, that connect two asymptotic
fourdimensional manifolds; they are interpreted as tunnelling between
the two asymptotic configurations. If a WH can be joined at $t=0$ to a
compact, hyperbolic universe, the euclidean solution can be interpreted
as nucleating a baby universe (BU) from the asymptotic region and gives
the semiclassical amplitude for creating a BU in the original space. The
BU then evolves according to its equations of motion.

A large amount of attention has been devoted to the explicit solution of
WH solutions. In particular, Giddings and Strominger [1] and Myers [2]
have discussed WHs generated from coupling the gravitational field to an
antisymmetric tensor of rank three (the axion), with topology $R\times
S^3$; Halliwell and Laflamme [3] have discussed solutions in presence of
a conformal massless field, and Coule and Maeda [4] have examined the
case of the axion field coupled to a scalar Klein--Gordon field (in both
cases with topology $R\times S^3$); Hawking [5] and Hosoya and Ogura [6]
have dealt with gravity coupled to a Yang - Mills field. The magnetic
monopole solution in four dimensions has been investigated by Keay and
Laflamme [7]; its topology is $R\times S^1\times S^2$.

In this paper we shall investigate a different WH solution of topology
$R\times S^1\times S^2$ generated by the electromagnetic (EM) field.

We shall first present the euclidean solution for the gravitational and
the EM fields in the case of zero and non zero cosmological constant;
then we discuss its analytic continuation at $t=0$ into a BU.

A different continuation in hyperbolic space leads to an alternative
interpretation: a Reissner--Nordstr\"om (RN)
type static solution with $M=0$
which is joined by the static wormhole to a second RN
space. We briefly discuss the behaviour of charged particles and their
crossing of the WH.

According to the usual interpretation, then, this is evidence for a
quantum tunnelling: the WH
yields the amplitude for transition from a RN space into a RN
isometric space. We discuss in detail the transition probability for
particles crossing between the two spaces.

This way of looking at the WH as a quantum bridge connecting two
classical
hyperbolic spaces opens the way to the interesting speculation that
singularities in the classical domain of physical, hyperbolic solutions
in general relativity can be avoided by euclidean solutions joining
two spaces, as it happens in the R--N case that we discuss here.

\beginsection 2. The euclidean solution.
Let us start from the euclidean action for gravity minimally coupled
to the EM field:
$$S_E=\int_\Omega d^4x\sqrt g\Biggl[-{M_p^2\over 16\pi G}(R+2\Lambda)+{1\over
4e^2}F^2\Biggr]+\int_{\partial\Omega} d^3x\sqrt h {{\bf K}\over 8\pi
G}.\eqno(2.1)$$
Here $\Omega$ is a compact fourdimensional manifold, $M_p$ is the Planck mass,
 $R$ is the curvature scalar, $\Lambda$ is the
cosmological constant, $F=F_{\mu\nu}F^{\mu\nu}$ the usual EM
lagrangian, ${\bf K}$ is the trace of the extrinsic curvature of the
boundary $\partial\Omega$ of $\Omega$
and $h$ is the determinant of the induced metric over $\partial\Omega$. For
$g_{\mu\nu}$ we look for a solution of the form
$$ds^2=dt^2+a^2(t)d\chi^2+b^2(t)d\Omega_2^2\eqno(2.2)$$
where $\chi$ is the coordinate of the 1-sphere, $0\le\chi<2\pi$ and
$d\Omega_2^2$ represents the line element of the 2-sphere. Let us first
discuss the case $\Lambda=0$. For the EM
field we choose the Ansatz
$$A_\mu=A(t)\delta_{\chi\mu}.\eqno(2.3)$$
The only non vanishing component of the EM field is of course
$$F_{t\chi}=-F_{\chi t}=\dot A(t).\eqno(2.4)$$
{}From the equation of motion
$$\partial_\mu(\sqrt g F^{\mu\nu})=0\eqno(2.5)$$
we obtain
$$\dot A=K{a\over b^2}\eqno(2.6)$$
($K$ is an integration constant).  Substituting (2.6) into (2.1) one
recovers after some algebra (details are given in the appendix) the scale
factors of the 1- and 2-sphere. The solution is
$$\eqalignno{ds^2&=dt^2+\bar c^2{t^2\over
c^2+t^2}d\chi^2+(c^2+t^2)d\Omega_2^2,&(2.7a),\cr
A(t)&=-{\bar cceM_p\over 2\sqrt\pi}{1\over\sqrt{c^2+t^2}}.&(2.7b)}$$
$c$ is connected to $K$ by
$$c^2={4\pi\over e^2}{K^2\over M_p^2};$$
$\bar c$ is an integration constant with dimension of length whose value
will remain arbitrary.

For $t^2\rightarrow\infty$, $a^2\rightarrow\bar c^2$ and
$b^2\rightarrow t^2$; these asymptotic behaviours ensure that the
solution can be interpreted as a WH connecting two asymptotic flat
space regions:
$$ds^2=dt^2+\bar c^2d\chi^2+t^2d\Omega_2^2.\eqno(2.8)$$
At $t=0$ the metric is singular. This is only due to the choice of the
coordinates, that cover only half of the WH. Indeed, in the
neighbourhood of $t=0$, redefining the variable $\chi$, the line element
becomes ($\bar c = c$)
$$ds^2=dt^2+t^2d\chi^2+c^2d\Omega_2^2.\eqno(2.9)$$
In the neighbourhood of $t=0$ our solution coincides with an euclidean
Kasner universe [8]. Naturally the singularity at $t=0$ can be
eliminated going to cartesian coordinates in the $(t,\>\chi)$ plane.
This particular case of singularity removable by a different choice of
coordinates has been classified by Gibbons and Hawking [9] as a `bolt'
singularity. In the neighbourhood of $t=0$ the topology is locally
$R^2\times S^2$ with $R^2$ contracting to zero as $t\rightarrow 0$. New
variables can be defined such that the whole euclidean space is
represented by a single chart. Let us define
$$t=c\tan{\xi\over 2}.\eqno(2.10)$$
$\xi$ is defined in the interval $(-\pi,\>\pi)$. Introduce the new
coordinates $u$, $v$ as
$$\eqalignno{u&={1-\cos{\xi\over 2}\over\sin{\xi\over
2}}e^{1/\cos{\xi\over 2}}\cos\chi,&(2.11a)\cr
v&={1-\cos{\xi\over 2}\over\sin{\xi\over 2}}e^{1/\cos{\xi\over
2}}\sin\chi.&(2.11b)\cr}$$
The expression of the line element shows that there is no singularity:
$$ds^2=c^2\biggl(1+{c \over \sqrt{t^2+c^2}}\biggr)^2
e^{-2\sqrt{t^2+c^2}/c}(du^2+dv^2)+(c^2+t^2)d\Omega_2^2.\eqno(2.12)$$
{}From
$$\eqalignno{u^2+v^2&={1-\cos{\xi\over 2}\over 1+\cos{\xi\over
2}}e^{2/\cos{\xi\over 2}},&(2.13a)\cr
{v\over u}&=\tan{\chi},&(2.13b)\cr}$$
we see that the geodesic at constant $\chi$ is a segment of a line
passing through the origin; geodesics at fixed $t$ are circumferences of
radius
$$r=\sqrt{{1-\cos{\xi\over 2}\over 1+\cos{\xi\over
2}}}e^{1/\cos{\xi\over 2}}.\eqno(2.14)$$

In section 5, eq.s (2.7) will be the starting point of the discussion of
the RN interpretation.

\beginsection 3. Non vanishing cosmological constant.
We start from (2.1). Using the same Ansatz (2.3) for the EM
field, a solution is now given by
$$\eqalignno{&ds^2={b^2\over\lambda b^4+b^2-c^2}db^2+\bar c^2{\lambda
b^4+b^2-c^2\over b^2}d\chi^2+b^2d\Omega_2^2,&(3.1a)\cr
&A(t)=-\bar c{K\over b},&(3.1b)\cr}$$
with $\lambda=\Lambda/3$. This solution reduces to (2.7) for vanishing
$\lambda$ with the substitution $b^2=c^2+t^2$.

Let us separately discuss $\lambda>0$ and  $\lambda<0$. In the first
case it is easy to see that eq. $(3.1a)$ is defined for
$$b^2>\tilde c^2={\sqrt{1+4\lambda c^2}-1\over 2\lambda}.\eqno(3.2)$$
Using the transformation
$$b^2=\tilde c^2+t^2\eqno(3.3)$$
eq. $(3.1a)$ takes the form
$$\eqalign{ds^2&={t^2\over\lambda(\tilde c^2+t^2)^2+\tilde
c^2+t^2-c^2}dt^2+\cr
&+\bar c^2 {\lambda(\tilde c^2+t^2)^2+\tilde
c^2+t^2-c^2\over\tilde
c^2+t^2}d\chi^2+(\tilde c^2+t^2)d\Omega_2^2\cr}\eqno(3.4)$$
where now $-\infty<t<+\infty$.

The asymptotic form of (3.4) for $t^2\rightarrow\infty$ is
$$ds^2={1\over\lambda t^2}dt^2+\lambda\bar c^2t^2d\chi^2+
t^2d\Omega_2^2.\eqno(3.5)$$
This is not a flat euclidean space. Let us redefine the euclidean time
by
$$\vert t\vert=\exp\bigl(\sqrt{\lambda t'^2}\bigr)\eqno(3.6)$$
so that
$$ds^2=dt'^2+e^{2\sqrt{\lambda t'^2}}\bigl(\lambda\bar c^2
d\chi^2+d\Omega_2^2\bigr).\eqno(3.7)$$
This line element defines an anisotropic universe whose scale factors
expand exponentially; their ratio is fixed by the value of the
cosmological constant.

For $\lambda<0$ $(3.1a)$ is defined when
$$\tilde c_a^2<b^2<\tilde c_b^2\eqno(3.8)$$
where
$$\eqalign{\tilde c^2_a&={1-\sqrt{1-4\vert\lambda\vert c^2}\over
2\vert\lambda\vert}\cr
\tilde c^2_b&={1+\sqrt{1-4\vert\lambda\vert c^2}\over
2\vert\lambda\vert}\cr}.\eqno(3.9)$$
and $(3.1a)$ can be interpreted like a tunnelling between two regions
that are the hyperbolic continuations of the metric $(3.1a)$ outside of
the range (3.9).

\beginsection 4. The solution in hyperbolic spacetime.
Half of our instanton can be joined to a real, hyperbolic signature
universe; this is the $bounce$ solution of the tunnelling.

Let us first investigate hyperbolic solutions of the coupled gravity and
EM field with the same symmetry as already investigated in the euclidean
case. The solution is given by
$$ds^2=-dt^2+\bar c^2{t^2\over
c^2-t^2}d\chi^2+(c^2-t^2)d\Omega_2^2.\eqno(4.1)$$

The line element (4.1) describes an anisotropic universe that lapses
$2c$ in time, born at $t=-c$ as a $spaghetti$ configuration; at $t=0$
contracts into a $pancake$ and again tends to $spaghetti$ for
$t\rightarrow c$. The EM field is imaginary, however the field strength
vanishes at $t=0$, so the joining is possible. It follows that the
universe $S^2\times S^1$ created at $t=0$ by tunnelling cannot propagate
in time by EM alone.

The problem is then which engine could power a BU obtained by tunnelling
described by half of our WH instanton solution. To this aim we can
exploit an interesting hyperbolic solution of the same symmetry, driven
by the axionic field. This solution has been obtained by Keay and
Laflamme [7] and makes use of the axionic field whose energy density is
negative in hyperbolic space.

Using the form (2.2) for the line element and introducing the axion
field
$$H_{\mu\nu\rho}=\epsilon_{t\mu\nu\rho}h(t)\eqno(4.2)$$
from the equation of motion one gets
$$h(t)={C\over ab^2}.\eqno(4.3)$$
where $C$ is a constant.
The solution for the line element has the form ($a$ is a constant with
dimension of length)
$$ds^2={b^2\over\lambda b^4+b^2-c'^2}db^2+a^2d\chi^2+b^2d\Omega_2^2
\eqno(4.4)$$
%
and
$$c'^2={{48\pi}\over{M_p^2}}C^2.$$
At $t=0$ and $\lambda=0$ eq. (4.4) becomes
$$ds^2_{(3)}=a^2d\chi^2+c'^2d\Omega_2^2.\eqno(4.5)$$
Let us define $y^2=t^2-\epsilon^2$ where $\epsilon$ is some
constant, hence using this into (2.7a) we obtain:
$$ds^2={y^2\over y^2+\epsilon^2}dy^2+\bar c^2{y^2+\epsilon^2\over
y^2+\epsilon^2+c^2}d\chi^2+(y^2+\epsilon^2+c^2)d\Omega_2^2.\eqno(4.6)$$
At $y=0$ we have
$$ds^2_{(3)}=\bar c^2{\epsilon^2\over\epsilon^2+c^2}d\chi^2+(\epsilon^2+c^2)
d\Omega_2^2.\eqno(4.7)$$
Identifying
$$\bar c^2{\epsilon^2\over\epsilon^2+c^2}=a^2$$
and
$$\epsilon^2+c^2=c'^2$$
we join our solution to the hyperbolic universe:
$$ds^2=-dt^2+\bar
c^2{\epsilon^2\over\epsilon^2+c^2}d\chi^2+(\epsilon^2+c^2-t^2)
d\Omega_2^2.\eqno(4.8)$$

In the general case $\lambda \not=0$ the euclidean line element (4.4)
can be interpreted as a tunnelling to a hyperbolic universe
$$ds^2=-{t^2\over t^2-\lambda(c^2-t^2)^2}dt^2+\bar c^2\lambda
c^2d\chi^2+(c^2-t^2)d\Omega_2^2.\eqno(4.9)$$

Let us now compute the amplitude for the formation of the BU in the case
$\Lambda=0$, starting from (2.1). Since $R=0$, the Euclidean action is
given by
$$S_E={1\over 4e^2}\int d^4 x \sqrt gF_{\mu\nu}F^{\mu\nu}.\eqno(4.10)$$
Remembering the form of the solution (2.7) we obtain for the nucleation
of the BU at $t=0$
$$\eqalign{S_E&={4\pi^2 K^2\over e^2}\int_0^{+\infty} dt{a\over b^2}\cr\cr
&=\pi M_p^2\bar cc\cr
&=2\pi^{3/2}\bar c{K\over e\sqrt G}.}\eqno(4.12)$$
The probability $\Gamma$ of formation of the BU in a Planck volume in a
Planck time is
$$\Gamma=e^{-S_E}=\exp{(-\pi M_p^2\bar cc)}.\eqno(4.13)$$
In order to have a finite probability, the order of magnitude of the
constants appearing in the solution is
$$\bar cc\approx {1\over \pi}l_p^2.\eqno(4.14)$$

\beginsection 5. The static wormhole interpretation.
The solution (2.7) can also be interpreted as an Euclidean
wormhole solution
joining two isometric, asymptotically flat space-times, described by a
RN type of solution.

To obtain this solution,
let us make a change of coordinates in (2.2) by substituting
$\chi\rightarrow iT$ and taking $T$ with dimension of length;
for clarity we shall also put $t\equiv r$. Throughout this section
we shall use geometrized units (velocity of light and gravitational
constant equal to one).
The solution we so obtain describes a hyperbolic space-time:
$$\eqalignno{ds^2&=dr^2-{r^2\over
r^2-Q^2}dT^2+(r^2-Q^2)d\Omega_2^2,&(5.1a)\cr
A(r)&=-{{K}\over{\sqrt{r^2-Q^2}}}.&(5.1b)}$$
where $Q$ is a constant.
The solution (5.1) is defined for $r^2>Q^2$. In this case,
denoting $R^2\equiv r^2-Q^2$, we obtain
$$ds^2
=-\biggl(1+{Q^2\over R^2}\biggr)dT^2+\biggl(1+{Q^2\over R^2}\biggr)^{-1}
dR^2+R^2d\Omega_2^2.\eqno(5.2)$$

This resembles a RN metric with source mass $M=0$ and
radial coordinate $R$ ranging now as $\infty>R\geq 0$.
At $R=0$ there is a true singularity.

In the region
$|r|<|Q|$ we have no hyperbolic solution, and
the overall picture is that of two branches of the
RN solution joined by the euclidean wormhole
solution (2.7) that is regular in this region. Formally, the solution
(5.1) can be obtained from (2.7) by $\bar c \chi\rightarrow iT$,
$c\rightarrow iQ$ since  the only non vanishing component of the
vector potential (2.3) is $A_{\chi}$.

The condition $M=0$ is rather peculiar since it implies that the
contribution to the gravitational mass of the source from the electric
charge is also zero [10,11],
hence the charge itself must be zero. This is consistent with
equation (2.5) which implies no charges and currents everywhere in the
domain of definition of the solution (5.2).

 The meaning of the parameter
$Q$ entering (5.2) requires some care. Since the electric field is
radial in $R$, its integral flux through a sphere containing the origin
$R=0$ is equal to
$$\Phi=4\pi Q\eqno(5.3).$$
The constant $Q$ then measures the electric field flux density through
the wormhole throat at $R=0$. Since there are no physical charges in the
field, the constant $Q$ only fixes, as boundary conditions for equation
(2.5), the amount of flux that we want through any given surface
containing the origin, similar to what is done for the axionic field in
[1]. Thus, since there is no support for charges and matter (hence
$M=0$), the electric field can extend smoothly beyond the wormhole
throat to the asymptotic infinity of the isometric space-time with
$R<0$, generating in both cases an apparent charge $Q$.

The traversability of the wormhole is at the basis of the
very usefulness of the wormhole concept; in order to cross the wormhole,
a classical particle must be able to reach it. Let us then study this
aspect of the solution introduced above.

Consider the equation of motion for a test particle, having
an electric charge per unit mass $q$ , total specific energy $E$ and
specific angular momentum $L$ with respect to the flat infinity, that
approaches $R=0$.
This is relevant since the particle can cross the
wormhole throat only if it gets $R=0$ (classically or via quantum
tunnelling). We assume the motion in
the equatorial plane, $\theta=\pi/2$.

The momenta and equations of motion are
$$\eqalignno{
P_T&\equiv -\biggl({R^2+Q^2\over R^2}\dot T+{\beta\over R}\biggr)=-E&(5.4a)\cr
P_R&\equiv{R^2\over R^2+Q^2}\dot R&(5.4b)\cr
P_\phi&\equiv R^2\dot\phi=L&(5.4c)\cr
\dot R^2&=\biggl(E-{\beta\over R}\biggr)^2-\biggl(1+{L\over
R^2}\biggr)\biggl(1+{Q^2\over R^2}\biggr)&{}\cr
&\equiv(E-V_+)(E-V_-)&(5.4d)\cr}$$
%
%
where $V_\pm$ are the  potential barriers given by [11]
$$V(R;\beta,{Q},L)_{\pm}=R^{-2}[\beta R\pm
(R^2+{Q}^2)^{1/2}(R^2+L)^{1/2}]\eqno(5.5)$$
with $\beta=qQ$. We shall study analytically the graph of the function
$V_+$. The behaviour of $V_-$ is easily deduced from the relation
$$V_-(\beta)=-V_+(-\beta).\eqno(5.6)$$
The analysis of the potential barriers (5.5) with $L\not=0$
shows that the barriers are repulsive for all values of the parameters.
On the contrary when $L=0$,
namely when the motion is strictly radial,
there is a class of trajectories which can
reach the wormhole throat ($R=0$). We shall discuss extensively this
latter case, referring to App. B for the general situation.

The potential barrier $V_+$ for $L=0$ reads, from (5.5):
$$V_{+}(R;\beta,Q)=R^{-1}\left[\beta+(R^2+Q^2)^{1/2}\right].\eqno(5.7)$$
When $R\to\infty$, $V_+$ behaves as
$$V_+\approx 1+{{\beta}\over{R}}$$
while when $R\to 0$ we have
$$V_+\approx{\beta+|Q|\over R}\to\cases{
\hbox{$+\infty$}&\hbox{$\beta>-|Q|$}\cr
\hbox{$0$}&\hbox{$\beta=-|Q|$}\cr
\hbox{$-\infty$}&\hbox{$\beta<-|Q|$}\cr}\eqno(5.8)$$

The conditions $V_+=1$ and $V_+=0$ are satisfied respectively when
$\beta=\beta_1\equiv R-(R^2+{Q}^2)^{1/2}$ and $\beta=\beta_0\equiv
-(R^2+{Q}^2)^{1/2}$. They are shown in figure 1a, while the graphs of
$V_+$ are shown in figure 1b for the cases $\beta>-|Q|$, $\beta=-|Q|$
and $\beta<-|Q|$.

We may repeat the analysis for the case of $V_-$ using the symmetry
(5.6). The conclusion is that the point $R=0$ can be reached if and only
if
$$\beta^2\geq Q^2.\eqno(5.9)$$
In this case the particles may reach the euclidean wormhole and
eventually emerge in the $R<0$ Universe [10, 12].
The transition probability $T_{WH}$ for tunnelling by the Euclidean
wormhole is proportional to $\exp(-2S_{\rm{cl}})$
where $S_{\rm{cl}}$, using (4.12) and remembering that $a$ is positive
(see (A.10) ) is given by
$$S_{cl}=\pi c\bar c M_p^2{\sqrt 2-1\over\sqrt 2}\eqno(5.10)$$

The transition probability characterizes the wormhole and is
independent of the particle's properties since the latters, provided
they satisfy (5.4), all reach the wormhole throat, regardless of
their energy.

The radial particles which do not satisfy (5.9) or those which
have a non zero angular momentum, may cross the wormhole reaching $R=0$
as a result of a quantum tunnelling with non zero quantum probability.
Indeed, let us
go back to eq. $(5.4d)$ and use it to establish the equation for the
wave function taking into account that $P_R\rightarrow -id/dR$ is given
by eq. $(5.4b)$:
$$-{{d^2}\over{dR^2}}\Psi~=~{{1}\over{(R^2+Q^2)^2}}~ \bigl[
R^2(RE-\beta)^2-(R^2+L)(R^2+Q^2)\bigr]\Psi\eqno(5.11).$$
The particle can reach $R=0$ by quantum barrier penetration; at $R=0$
the solution is joined to the euclidean WH connecting to the second
region ($R<0$ for convenience). The overall transition probability is
given by
$$T~=~T_0^2~T_{WH}\eqno(5.12)$$
where $T_{WH}$ is due to the tunnelling by the euclidean wormhole
and $T_0$ is the usual quantum transition probability of the barrier
from $R=R_0$ to $R=0$.

For the evaluation of $T_0$ in the WKB approximation the
relevant quantity is
$$\alpha=\int_0^{R_0}dR{1\over {R^2+Q^2}}\bigl[(R^2+L)(R^2+Q^2)-
R^2(ER-\beta)^2\bigr]^{1/2}\eqno(5.13)$$
which is finite and has a particularly simple expression for $L=0$.

In conclusion, in the present section we have used the euclidean
solution in order to obtain a finite traversability amplitude
from a RN space-time into an isometric space-time;
we may call this a space-tunnelling WH. This also yields
an  interpretation of the $M=0$ RN space-time.

Our interpretation of a space-tunnelling WH is in the direction
of the proposal
by John Wheeler [13; see also 14] of a pair of oppositely charged
extreme RN black holes identified at their throats, in an
electromagnetic field. In the  present case the joining of two
RN space-times happens through quantum tunnelling.

\beginack
We are glad to thank Paolo Ciatti for interesting discussions.

\beginappendix A.
The Einstein equations are
$$R_{\mu\nu}-{1\over 2}g_{\mu\nu}R=8\pi GT_{\mu\nu}.\eqno(A.1)$$
Substituting eqs. (2.4) in the expression for $T_{\mu\nu}$ we may find
its components in terms of the scale factors:
$$\eqalign{T_{tt}&={1\over 2e^2}{K^2\over b^4},\cr
T_{\chi\chi}&={1\over 2e^2}{K^2\over b^4}a^2,\cr
T_{ij}&=-{1\over 2e^2}{K^2\over b^4}g_{ij},\cr}\eqno(A.2)$$
The ensuing equations for the two scale factors are then
$$\eqalignno{{\dot b^2\over b^2}-{1\over b^2}+2{\dot a\dot b\over
ab}&={c^2\over b^4},&(A.3a)\cr
2{\ddot b\over b}+{\dot b^2\over b^2}-{1\over b^2}&={c^2\over
b^4},&(A.3b)\cr
{\ddot b\over b}+{\ddot a\over a}+{\dot a\dot b\over ab}&=-{c^2\over
b^4}.&(A.3c)\cr}$$
In $(A.3b)$ only $b(t)$ appears; by the substitution $\dot b=p$,
$(A.3b)$ takes the form
$$f'+{1\over b}f-{1\over b}={c^2\over b^3}\eqno(A.4)$$
where $f=p^2$ and we have used $\ddot b=pp'$. Putting then $b=e^h$ we
get
$${df\over dh}+f=1-c^2e^{-2h}\eqno(A.5)$$
whose general solution is
$$f=K_1e^{-h}+1-c^2e^{-2h}.\eqno(A.6)$$
$K_1$ is an integration constant. In what follows we shall simply take
$K_1=0$. In the old variables then
$$\dot b^2=1-{c^2\over b^2}.\eqno(A.7)$$
So finally
$$b(t)=\sqrt{c^2+t^2}.\eqno(A.8)$$
By substitution of (A.8) into $(A.3a)$ the equation for the remaining
scale factor is
$${\dot a\over a}={c^2\over t(c^2+t^2)}\eqno(A.9)$$
whose solution is
$$a(t)=\pm\bar c{t\over\sqrt{c^2+t^2}}\eqno(A.10)$$
where $\bar c$ is an integration constant.

The signs $\pm$ refer respectively to the submanifolds with $t>0$ or
$t<0$, having defined $a(t)$ as non negative.

Let us list for completeness the equations corresponding to (A.3) in the
case of hyperbolic signature:
$$\eqalignno{{\dot b^2\over b^2}+{1\over b^2}+2{\dot a\dot b\over
ab}&=-{c^2\over b^4},&(A.12a)\cr
2{\ddot b\over b}+{\dot b^2\over b^2}+{1\over b^2}&=-{c^2\over
b^4},&(A.12b)\cr
{\ddot b\over b}+{\ddot a\over a}+{\dot a\dot b\over ab}&=+{c^2\over
b^4},&(A.12c)\cr}$$
from which it is easy to obtain the line element (4.1)

{}From $(A.12a)$ we may control that $T_{tt}$ is negative:
$$T_{tt}=-{1\over 2e^2}{K^2\over b^4}.\eqno(A.13)$$
The electromagnetic field is imaginary; thus this solution does not
describe a real evolution in hyperbolic space.

\beginappendix B.
We give here the details of the equations of motion in the case
$L\not=0$. The asymptotic behaviour of $V_+$ for large $R$ is
$$V\approx{{\beta+R}\over{R}}\to\cases{
\hbox{$1_+$}&\hbox{$\beta\geq 0$}\cr
\hbox{$1_-$}&\hbox{$\beta<0$}\cr}$$
For $R\rightarrow 0$ we have
$$V_+\approx{{|Q|\sqrt{L}}\over{R^2}}\to +\infty$$
The condition $V_+=1$ is equivalent to
$$\beta=R^{-1}\left[R^2-(R^2+Q^2)^{1/2}(R^2+L)^{1/2}\right]
\equiv\beta_1.\eqno(B.1)$$
Clearly $\beta_1<0$ always and $\lim_{R\to\infty}\beta_1=0$, $\lim_{R\to
0}\beta_1=-\infty$. The function $\beta_1$ is plotted in figure 2b. The
condition $V_+=0$ is satisfied when
$$\beta=-R^{-1}~(R^2+Q^2)^{1/2}(R^2+L)^{1/2}\equiv\beta_0.
\eqno(B.2)$$
Here again $\beta_0<0$ always; the graph of $\beta_0(R)$ is easily
deduced from its limits
$$\lim_{R\to\infty}\beta_0=\lim_{R\to 0}\beta_0=-\infty$$
and from the locus of its critical points, namely
$$L=R^4{Q}^{-2}\equiv L_c.$$
The function $L_c$ is plotted in figure 2a in the ($L-R$)-plane. From
(B.1) and (B.2) we find $\beta_1=R+\beta_0$ hence $\beta_1\geq\beta_0$,
the equality sign holding only in the limit $R\to 0$. The value of
$\beta_0$ at its maximum is given by
$$\beta_0(R;Q,\,L_c)\equiv\beta_{0c}=-{{R^2+Q^2}\over{|Q|}}$$
and its graph is plotted in figure 2b (dashed line). We are now in the
position to draw the potential curves $V_+(R;\,\beta, Q,L)$ as function
of $R$ for any given set of values ($\beta,Q,L$). They are shown in
figure 2c for three different values of $\beta$, namely 1) $\beta>0$; 2)
$\beta<0$ and $\beta>\beta_{0c}$; 3) $\beta<0$ and $\beta<-\beta_{0c}$.

The classical motion is only allowed when the total energy $E$ of the
charged test particle satisfies the condition $E\geq V$, hence, when the
angular momentum $L$ is different from zero, we see by a direct
inspection of figure 2 that,

i) the wormhole throat $R=0$ cannot be reached classically since the
field is repulsive to all particles, either charged or not;

ii) a sea of negative energy particles is allowed in the vicinity of the
throat. This effect is a well known property of the
RN solution and allows for electric field energy
extraction via quantum tunnelling.

\vfill\eject

\beginref
\ref [1] Giddings, S. B. and A. Strominger, 1988, {\tscors
Nucl. Phys.} B {\bf 306}, 890.
\ref [2] Myers, R. C., 1988, {\tscors Phys. Rev.} D {\bf 38}, 1327.
\ref [3] Halliwell, J. J. and R. Laflamme, 1989, {\tscors Class.
Quantum Grav.} {\bf 6}, 1839.
\ref [4] Coule, D. H. and K. Maeda, 1990, {\tscors Class. Quantum
Grav.} {\bf 7}, 955.
\ref [5] Hawking, S. W., 1987, {\tscors Phys.  Lett.} B {\bf 195}, 337.
\ref [6] Hosoya, A. and W. Ogura, 1989, {\tscors Phys. Lett.} B {\bf
225}, 117.
\ref [7] Keay, B. J. and R. Laflamme, 1989, {\tscors Phys. Rev.} D
{\bf 40}, 2118.
\ref [8] Misner C. W., K. S. Thorne and J. A. Wheeler, {\tscors
Gravitation}, W. H. Freeman and Company, New York 1973.
\ref [9] Gibbons, G. W. and S. W. Hawking, 1979, {\tscors Commun. Math.
Phys.} {\bf 66}, 291.
\ref [10] Cohen J.M. and R. Gautreau, 1979, {\tscors Phys. Rev D}
{\bf 19}, 2273.
\ref [11] de Felice F. and C.J.S. Clarke, 1990, {\tscors Relativity on
Curved Manifolds,} Cambridge University Press, Cambridge, England.
\ref [12] Graves J.C. and  D.R. Brill, 1960, {\tscors Phys. Rev.}
{\bf 120}, 1507.
\ref [13] Wheeler J.A., 1962 {\tscors Geometrodynamics,} Academic,
New York.
\ref [14] Garfinkle D. and A. Strominger, 1991, {\tscors Phys. Lett. B}
{\bf 256}, 146.

\endref
\vfill\eject

\begincaptions
Fig. 1 - a) Plot of the functions $\beta_1$ (solid line)
and $\beta_0$ (dashed line) when $L=0$. b)
Behaviour of the effective potential $V_+$ when $\beta>-|Q|$ (solid
line);  $\beta=-|Q|$ (dashed); $\beta<-|Q|$ (dot-dashed).
\medskip
Fig. 2 - a) Plot of the function $L_c$ which is the locus of the points
where the function $\beta_0$ has a maximum. b) Plots of the function
$\beta_1$ (solid line) which identifies where $V_+=1$
and of the function $\beta_0$ (dot-dashed)
which identifies where $V_+=0$. The locus of the maxima of $\beta_0$ is
along the dashed curve, plot of the function $\beta_{0c}$; c) behaviour
of the effective potential $V_+$ as function of $R$.
The solid line represents the case
$\beta >0$; dashed line for $\beta<0$ and $\beta>\beta_{0c}$;
dot-dashed when $\beta<0$ and $\beta<\beta_{0c}$.
\vfill\eject
\bye